\documentclass[12pt]{article}

\usepackage[titletoc,title]{appendix}
\usepackage[latin1]{inputenc}
\usepackage{amsmath}
\usepackage{amssymb}
\usepackage{graphicx}
\usepackage[margin=1.25in]{geometry}
\usepackage[bottom]{footmisc}
\usepackage{indentfirst}
\usepackage{endnotes}
\usepackage{mathabx}
\usepackage{enumerate}
\usepackage{rotating}
\usepackage{multirow}
\usepackage{booktabs,calc}
\usepackage{threeparttable}
\usepackage{threeparttable}
\usepackage{float}
\usepackage{url}
\usepackage{epsfig}
\usepackage{caption,subcaption}
\usepackage{float}
\usepackage{url}
\usepackage{amsmath,amsfonts}
\usepackage{amsthm}
\usepackage{mathrsfs}

\usepackage[onehalfspacing]{setspace}
\usepackage{apptools}

\newcommand{\suchthat}{\;\ifnum\currentgrouptype=16 \middle\fi|\;}

\newtheorem{assm}{Assumption}

\newtheorem{prop}{Proposition}
\newtheorem{lemma}{Lemma}

\theoremstyle{remark}

\usepackage{chngcntr}
\usepackage{graphicx}
\usepackage{amsmath}
\usepackage{amssymb}
\usepackage{parskip}
\usepackage{sgame}
\usepackage{color}
\usepackage{tikz}
\usetikzlibrary{trees,calc}

\usepackage{natbib}

\begin{document}

\title{\large{Obstacles to Redistribution Through Markets and One Solution}\thanks{We thank Piotr Dworczak, Bruno Salcedo, and Charles Zheng for helpful comments. Any remaining errors are our own.}}

\author{
    \small{Roy Allen}  \\
    \small{Department of Economics} \\
    \small{University of Western Ontario} \\
    \small{rallen46@uwo.ca}\\
    \and
    \small{John Rehbeck}  \\
    \small{Department of Economics} \\
    \small{The Ohio State University} \\
    \small{rehbeck.7@osu.edu}\\
}
\date{\small{ \today }}

\maketitle
\begin{abstract}
\cite{dworczak2021redistribution} study when certain market structures are optimal in the presence of heterogeneous preferences. A key assumption is that the social planner knows the joint distribution of the value of the good and marginal value of money. This paper studies whether relevant features of this distribution are identified from choice data. We show that the features of the distribution needed to characterize optimal market structure cannot be identified when demand is known for all prices. While this is a negative result, we show that the distribution of good value and marginal utility of money is fully identified when there is an observed measure of quality that varies. Thus, while \cite{dworczak2021redistribution} abstract from quality, we show how including quality is crucial for potential applications.\\

\end{abstract}

\newpage
\setlength\parindent{24pt}
\section{Introduction}
\cite{dworczak2021redistribution} asks several important economic questions such as: When are price regulations in a market optimal? What is the structure of optimal price regulation? Can redistributive policies improve social welfare? \cite{dworczak2021redistribution} elegantly provides answers to each question. The main primitives needed to characterize the optimal market structure are certain features of the joint distribution of the value of the good and marginal value of money. In their analysis, these are assumed known by the social planner \emph{a priori}.\footnote{In this paper, we focus only on identification of the demand side since the supply side has the same underlying structure in \cite{dworczak2021redistribution}.} Thus, to apply these results in practice, it is crucial to know whether relevant features of the joint distribution on preferences can be identified. This paper studies identification of this joint distribution from choice data.

We present two results. First, even when demand is well-behaved and observed without error for all prices, this is not enough to recover the features needed to characterize optimal market structure. While this is a negative result, our second result shows that we can identify the entire joint distribution of preferences when there is an observed measure of the quality of the good. Thus, while \cite{dworczak2021redistribution} abstract from the quality of goods present in real-world markets, we show that quality information is crucial for the planner to learn the distribution of preferences. Fortunately, such quality data is present for all markets mentioned in \cite{dworczak2021redistribution}. In the Iranian kidney market, the quality of kidneys can be measured by individual health histories. In the rental real estate market, apartments differ on various measures of quality such as location or square footage. Finally, in the labor market, jobs that have similar pay may have different company culture or benefits. 

\section{Model}\label{sec:model}
We consider the framework of \cite{dworczak2021redistribution} and examine when choice data can or cannot identify the joint distribution of good value and marginal value of money. Understanding this feature of the model is crucial to apply the results from \cite{dworczak2021redistribution} since the conditions that characterize the optimal mechanism depend on features of this distribution. We briefly summarize the model for the non-owners of the good (demand side) from \cite{dworczak2021redistribution}. 

Here there exists a unit mass of \emph{non-owners} who have preferences for a good ($K$) and money ($M$). The demand side is assumed to have no units of the good $K$ and have unit demand for the good. Each individual has a value for the good ($v^K$) and a value for money ($v^M$). Let $(x^K,x^M) \in \{0,1\} \times \mathbb{R}$ where $x^K$ describes whether the individual purchases the good and $x^M$ is the amount of money the person holds. 

Each individual is assumed to receive utility  
\[ v^K x^K + v^M x^M.\]
The value vector $(v^K,v^M)$ is assumed to be distributed according to a joint distribution $F(v^K,v^M)$. For simplicity, we suppose that the joint distribution has a probability density function given by $f(v^K,v^M)$. We also suppose that $v^K$ and $v^M$ have non-negative and bounded support. These are stronger assumptions than \cite{dworczak2021redistribution}, but seem reasonable for application. Moreover, since our first result is a non-identification result, we show that even under additional structure, we cannot recover relevant features of the distribution.

In the framework of \cite{dworczak2021redistribution}, the characterization of the optimal mechanism depends on the magnitude of certain moments of the distribution $F(v^K,v^M)$. In particular, \cite{dworczak2021redistribution} shows that the optimal market structure depends on the values 
\[ \mathbb{E}[v^M] \quad \text{and} \quad \mathbb{E}[v^M \mid \underline{r}] \] 
where $\underline{r}$ is the minimum of $\frac{v^K}{v^M}$ over its support. Similarly, $\overline{r}$ is the maximum of $\frac{v^K}{v^M}$ over its support. Many of the propositions depend on whether there is low or high inequality of valuations. Inequality on the demand side of the market is said to be low when $\mathbb{E}[v^M \mid \underline{r}] \le 2\mathbb{E}[v^M]$ and high when the opposite strict inequality holds.  

\section{Non-identification with Demand}\label{sec:nonidentification}
We first focus on what can be identified with demand data. We show that even when the demand function is known ex-ante, we cannot recover the features of the distribution needed to characterize the optimal mechanism. Let $p \in \mathbb{R}_+$ be the price of the good. 

When individuals have preferences from Section~\ref{sec:model}, each individual chooses to buy one unit of good $K$ when $\frac{v^K}{v^M} > p$, regardless of their initial monetary holdings. Thus, demand is given by $D(p)=1-\int \int 1 \left\{ \frac{v^K}{v^M} \le p \right\} f(v^K,v^M) dv^K dv^M$ where $1\{\cdot \}$ is an indicator function. 

We note that knowledge of $F(v^K,v^M)$ is equivalent to knowing the joint distribution over $r$ and $v^M$ given by $G(r,v^M)$ where $r = \frac{v^K}{v^M}$, since this gives a one-to-one transformation of random variables under the maintained assumption that $v^M \neq 0$ almost surely. We denote the  joint probability density function of $G(r,v^M)$ by $g(r,v^M)$, which exists since our distribution is continuous and we have an almost everywhere non-zero Jacobian for the change of variables.

As in \cite{dworczak2021redistribution}, it will be useful to reference the marginal cumulative density function of  $r=\frac{v^K}{v^M}$. Here, we abuse notation so that the cumulative distribution over $r$ is given by  $G(r)=\int \int 1\{ \frac{v^K}{v^M} \le r \} f(v^K, v^M) d v^K d v^M$. Similarly, we denote the probability density function over $r$ as  $g(r)$. Thus, it follows that knowing the market demand for good $K$ is equivalent to knowing the cumulative density distribution of $r$ since
\[ D(p) = 1-G(p).\] 
Since demand and the marginal distribution of $r$ are directly related, knowledge of demand gives $G(r)$, which we record below.

\begin{lemma} 
If $D(p)$ is known for all prices, then $G(r)$ is known. 
\end{lemma}

Under the assumptions on the joint distribution $F(v^K,v^M)$, the demand function will satisfy several nice properties. For example, demand is monotone decreasing in price. Also, we know that $D(p)=1$ for all $p \le \underline{p}$ and $D(p)=0$ for all $p \ge \overline{p}$ where $\underline{r}=\underline{p}$ and $\overline{r}=\overline{p}$. Since $G(p)$ has a probability density function, we know that $D(p)$ is differentiable on $(\underline{p},\overline{p})$.

We will show that when demand is known, one cannot identify whether same the demand side inequality is low or high. The reason for this is that demand only gives us information on the marginal distribution of $G(r,v^M)$ with respect to $r$ but does not restrict conditional moments of $v^M$. To see this, assuming that conditional moments exist and using the law of iterated expectations, we see that
\begin{align*}
    \mathbb{E}[v^M] &= \int \int v^M g(r,v^M) dv^M dr \\
            &= \int \mathbb{E}[v^M \mid r] g(r) dr.
\end{align*}
Thus, even though $g(r)$ is identified from demand, there is limited information about the conditional expectation of $v^M$ given $r$. We show in the proof of the following proposition that even when demand is known there can be low or high  inequality on the demand side of the market.

\begin{prop}
Knowledge of the demand $D(p)$ does not identify whether there is low or high same side inequality of demand.
\end{prop}

\begin{proof}
To see this, note there are no restrictions on the conditional demand. Therefore, we look for an arbitrary function $h(r)$ that maps to finite non-negative numbers to get
\[ \mathbb{E}[v^M \mid r] = \frac{h(r)}{g(r)} \]
where $g(r)$ is the probability density distribution identified from demand. Note that this expectation can be generated by letting the distribution of $v^M$ conditional on $r$ be a truncated normal distribution with mean $\frac{h(r)}{g(r)}$ and truncated at $\frac{h(r)}{g(r)}-\varepsilon$ and $\frac{h(r)}{g(r)}+\varepsilon$ where $\varepsilon \in \left(0, \frac{h(r)}{g(r)}\right)$, so the marginal value of money is non-negative.

Note that when the conditional mean satisfies
\[ \mathbb{E}[v^M \mid r] = \frac{h(r)}{g(r)} \]
it follows that 
\[ \mathbb{E}[v^M] = \int_{\underline{r}}^{\overline{r}} h(r) dr\]
where we are essentially finding the expectation of $h(r)$ over a uniform distribution. Thus, we see that the condition to determine market inequality reduces to
\begin{equation}\label{eq:ineq} h(\underline{r}) \gtreqless 2 g(\underline{r}) \int_{\underline{r}}^{\overline{r}} h(r) dr \end{equation}

For example, inequality is low when the scaled area under $h(r)$ exceeds $h\left(\underline{r}\right)$ as stated below
\begin{equation}\label{eq:low}
h(\underline{r}) \le 2 g(\underline{r}) \int_{\underline{r}}^{\overline{r}} h(r) dr. \end{equation}
We show one function that gives low market inequality. Let $h(r) = r - \underline{r} + \delta$ where $\delta>0$ so that
integrating Equation~\ref{eq:low} gives
\[ \delta \le 2 g(\underline{r}) \int_{\underline{r}}^{\overline{r}} (r-\underline{r}) dr + 2 \delta g(\underline{r}) (\overline{r}-\underline{r}) \] 
which is true when $\delta \le 2 g(\underline{r}) \int_{\underline{r}}^{\overline{r}} (r-\underline{r}) dr$. 

Similarly, high same side inequality of demand occurs when 
\begin{equation}\label{eq:high} h(\underline{r}) > 2 g(\underline{r}) \int_{\underline{r}}^{\overline{r}} h(r) dr. \end{equation}
One function that gives high same-side inequality is $h(r) = \frac{1}{\sqrt{r-\underline{r}+\delta}}$. Integrating Equation~\ref{eq:high} gives
\[\frac{1}{\sqrt{\delta}} > 4 g(\underline{r}) \left( \sqrt{\overline{r}-\underline{r}+\delta} - \sqrt{\delta} \right) \]
which is implied when
\begin{equation}\label{eq:quad}
\delta^2+(\overline{r}-\underline{r})\delta - \frac{1}{16g^2(\underline{r})} < 0.
\end{equation}
Finding the roots of this, we find that Equation~\ref{eq:quad} holds when\\ $\delta < \frac{1}{2} \left( \sqrt{(\overline{r}-\underline{r})^2 + \frac{1}{4g^2(\underline{r})}} - (\overline{r}-\underline{r})\right)$.

This shows that information about $\mathbb{E}[v^M]$ and $\mathbb{E}[v^M \mid \underline{r}]$ needed to determine low or high inequality cannot be recovered from demand data.
\end{proof}

\section{Identification with Quality}\label{sec:id}

The previous section shows how demand data with only prices cannot identify the key features needed to characterize the optimal mechanism from  \cite{dworczak2021redistribution}. This section shows that if the analyst also observes variation in \textit{quality} of the good, then the previous non-identification result is overturned.\footnote{The way we introduce quality differs from \cite{akbarpour2020redistributive} who look at optimal allocation policies that depend on  continuous good quality with transfers.}  To formalize this, we consider buyers with preferences over $(x^K, x^Q, x^M) \in \{ 0, 1 \} \times \mathbb{R}^2$, where $x^Q$ is a measure of quality. Here we imagine quality is constant within a sub-market, but varies exogenous across sub-markets. This is a common assumption in applied work. 

Now, the utility of individuals on the demand side is given by
\[ 
(v^K + v^Q x^Q) x^K + v^M x^M.
\]
Here, $v^Q$ is the value of quality. When quality is fixed in a market, this is exactly the setup of \cite{dworczak2021redistribution} once we define $\tilde{v}^K = v^K + v^Q x^Q$ as the value of the good. We will provide conditions under which the joint distribution of $(v^K, v^Q, v^M)$ is identified. This in turn identifies the joint distribution of $(\tilde{v}^K, v^M)$. Thus, the analysis of \cite{dworczak2021redistribution} can be used in markets where the quality $x^Q$ is the same.

We assume that the distribution of $(v^K, v^Q, v^M)$ admits a density of the form $f(v^K, v^Q, v^M)$. For this situation, the demand curve depends on quality $x^Q$ and price $p$. We write demand with quality as 
\[
D_Q(x^Q,p) = \int \int \int 1 \{ v^K + v^Q x^Q - v^M p \geq 0 \} f(v^K, v^Q, v^M) d v^K d v^Q d v^M.
\]
This specification uses the fact that the individual pays price $p$ for the good and thus faces the dis-utility of expenditure $-v^M p$ regardless of monetary holdings. We give assumptions below that ensure identification of the distribution of values.

\begin{assm} \label{assm:set1a}
The quality-price demand function $D_Q$ is known for all values $x^Q \in \mathbb{R}$ and $p \in B \subseteq \mathbb{R}_{+}$, where $B$ contains an open set.
\end{assm}

\begin{assm} \label{assm:set1b}
The value $v^Q = 1$ almost surely, the distribution of $(v^K, v^M)$ is determined by its moments, and all absolute moments of $(v^K, v^M)$ exist and are finite.
\end{assm}

The assumption $v^Q = 1$ sets the scale of the latent utility model. Some assumption on scale is necessary to identify the distribution of $(v^K, v^M)$. When $v^Q > 0$ almost surely, the extra assumption $v^Q = 1$ almost surely does not have any additional empirical restrictions, and is thus a normalization.

\begin{prop}
Under Assumptions~\ref{assm:set1a} and~\ref{assm:set1b}, the distribution of $(v^K, v^M)$ is identified.
\end{prop}

\begin{proof} With the assumption $v^Q = 1$ almost surely, write
\[
D_Q(x^Q,p) = \int \int \int 1 \{ v^K - v^M p \geq -x^Q \} f(v^K, v^Q, v^M) d v^K d v^Q d v^M.
\]
We recognize that varying $x^Q$ over all of $\mathbb{R}$ recovers the distribution of $v^K - v^M p$. By varying $p$ in an open ball contained in $B$, we recover the distribution of $(v^K,v^M)$ from \cite{masten2018random}, Lemma 2.
\end{proof}

\cite{fox2021note} presents a related identification result for discrete choice.

\bibliographystyle{ecta}
\bibliography{ref}

\end{document}